\documentclass[letterpaper]{article} 
\usepackage{aaai24}  
\usepackage{times}  
\usepackage{helvet}  
\usepackage{courier}  
\usepackage[hyphens]{url}  
\usepackage{graphicx} 
\usepackage{natbib}  
\usepackage{caption} 
\usepackage[ruled,vlined,linesnumbered]{algorithm2e}
\usepackage{algorithmic}
\usepackage{bibentry}
\usepackage{amsmath} 
\usepackage{bm}
\usepackage{booktabs}
\usepackage{subfigure}
\usepackage{amsmath,amssymb,amsfonts}
\usepackage{multirow}
\usepackage{graphicx}
\usepackage{textcomp}
\usepackage{xcolor}
\usepackage{amsthm}
\newtheorem{theorem}{Theorem}

\newtheorem{assumption}{Assumption}
\newcommand{\methodname}{{\tt{FedAIS}}}

\usepackage{newfloat}
\usepackage{listings}
\DeclareCaptionStyle{ruled}{labelfont=normalfont,labelsep=colon,strut=off} 
\nocopyright 

\title{Federated Graph Learning with Adaptive Importance-based
Sampling}

\author{Anran Li$^1$, Yuanyuan Chen$^2$, Chao Ren$^2$, Wenhan Wang$^3$, Ming Hu$^4$, Tianlin Li$^2$, \\ Han Yu$^2$, Qingyu Chen$^1$  
}
\affiliations{
$^1$ Department of Biomedical Informatics \& Data Science, School of Medicine at Yale University; 
$^2$School of Computer Science and Engineering, Nanyang Technological University, Singapore; 
$^3$ University of Alberta; \\ 
$^4$ School of
Computer Science and Engineering, Singapore Management University 
}

\begin{document}

\maketitle

\begin{abstract}
For privacy-preserving graph learning tasks involving distributed graph datasets, federated learning (FL)-based GCN (FedGCN) training is required. 
A key challenge for FedGCN is scaling to large-scale graphs, which typically incurs high computation and communication costs when dealing with the explosively increasing number of neighbors. 
Existing graph sampling-enhanced FedGCN training approaches ignore graph structural information or dynamics of optimization, resulting in high variance and inaccurate node embeddings. 
To address this limitation, we propose the \underline{Fed}erated \underline{A}daptive \underline{I}mportance-based \underline{S}ampling (\methodname{}) approach. It achieves substantial computational cost saving by focusing the limited resources on training important nodes, while reducing communication overhead via adaptive historical embedding synchronization. 
The proposed adaptive importance-based sampling method jointly considers the graph structural heterogeneity  and the optimization dynamics to achieve optimal trade-off between efficiency and accuracy.
Extensive evaluations against five state-of-the-art baselines on five real-world graph datasets show that \methodname{} achieves comparable or up to 3.23\%  higher test accuracy, while saving communication and computation costs by 91.77\% and 85.59\%. 


\end{abstract}

\section{Introduction}
Graph convolutional networks (GCNs) \cite{kipf2016semi, fey2021gnnautoscale,chen2018fastgcn} have achieved impressing performance for a wide range of learning tasks on graph data. 
However, due to privacy concerns,  heterogeneous subgraphs are separately stored by different data owners, constructing globally applicable GCNs requires collaborative learning. 
Federated graph learning (FedGL) for collaborative GCN training while preserving data privacy has attracted increasing attention \cite{he2021fedgraphnn,chen2021fedgraph,liu2022federated}. 
Based on the distribution of graph data, FedGL can be divided into inter-graph FedGL \cite{he2021fedgraphnn} and intra-graph FedGL \cite{chen2021fedgraph}. 
Intra-graph FedGL is common in practice \emph{e.g.}, in an online social application where each user has a local social network which contains interests and user interactions, and all networks form the latent complete human social network. 


However, training intra-graph FedGL on large-scale heterogeneous graphs remains a challenge. 
Firstly, the exponentially increasing dependency of neighbor nodes over layers (\emph{i.e.}, neighbor explosion) causes the computation graph to be extremely large, which incurs prohibitively high computation cost. 
Secondly, intra-graph FedGL requires transferring  intermediate embeddings across clients. This involves large number of edges connecting nodes that are stored by different clients. 
Since calculating embeddings for a node requires embeddings from its recursive neighbors several hops away, which could be stored by other clients, fetching neighbor embeddings across clients incurs high communication cost. 
Ignoring the information from neighbors across clients and treating subgraphs at various clients as independent can degrade model performance \cite{chen2021fedgraph}. 
Thirdly, client data in intra-graph FedGL exhibit statistical and structural heterogeneity. Different subsets of training data leads to different model accuracy and latency. 


Existing methods of efficient FedGCN training can be categorized into three categories. The first category uses missing neighbor generation \cite{zhang2021subgraph, zhang2021fastgnn} to acquire accurate node embeddings. However, the additional training of the generative models would incur high computation and communication costs. 
The second category focuses on graph sampling \cite{zhang2021subgraph}, \emph{e.g.}, FedGraph \cite{chen2021fedgraph} uses deep reinforcement learning (DRL) to select neighbor nodes for embedding aggregation. 
However, it ignores the graph structural heterogeneity information, resulting in inaccurate node embeddings and large variance in the gradients. 
The third category periodically transfer cross-client neighbor embedding transmission \cite{chen2021fedgraph, zhang2022fedego, deng2023graphfed} to reduce communication costs. However, it fails to capture the dynamics of model training to determine the optimal transfer period, resulting in inferior model performance. 
As for graph sampling for centralized learning, there are three types: 1) \textit{node-wise sampling} methods iteratively sample a number of neighbor nodes for each node \cite{hamilton2017inductive, dai2018learning}; 2) \textit{layer-wise sampling} methods select a number of nodes for each GCN layer \cite{chen2018fastgcn, zou2019layer}; and 3) \textit{subgraph sampling} methods sample a number of subgraphs from each training batch \cite{chiang2019cluster, zeng2019graphsaint}. 
However, since these approaches are not designed for FL, they require access to potentially sensitive raw data which risk privacy of local clients. Besides, such methods neglect communication costs and would incur substantial communication burden when the volume of the graph data is large. 


 

%

To address these limitations, we propose a novel federated graph sampling scheme - the \underline{Fed}erated \underline{A}daptive \underline{I}mportance-based \underline{S}ampling (FedAIS) approach for large-scale graph data node classification tasks.
It reduces the graph sampling variance and achieves substantial cost savings by efficiently leveraging historical embedding estimators and focuses the limited communication and computation resources on training important local samples. 
By designing an adaptive embedding synchronization scheme, it is capable of achieving the optimal trade-off between test accuracy and computation and communication cost savings. 
The key advantages of \methodname{} are summarized as follows.  
\begin{itemize}
    \item \textbf{Scalability:} 
    \methodname{} is able to scale FedGCN to large graphs with a constant memory cost with respect to input node sizes. For a selected set of batches, \methodname{} prunes the GCN computation graph so that only nodes inside the current batches and their direct 1-hop cross-client neighbors are retained, regardless of the depth of the GCN. Historical embeddings are used to accurately fill in the inter-dependency information of cross-client neighbors. 
    \item \textbf{Efficiency:} \methodname{} achieves highly efficient FedGL 
    and reduces unnecessary sample training via dynamic importance-based sampling that considers both structural information and optimization dynamics. It reduces cross-client neighbor embedding communication through adaptive embedding synchronization to select the optimal transmission interval that achieves the fast decay of the objective function. 
    \item \textbf{Convergence:} \methodname{} ensures that the global model converges in an efficient manner. Theoretical analyses show that the approximation variance induced by importance-based node sampling and the staleness of historical embedding is upper bounded. 
\end{itemize}

We evaluate \methodname{} on five graph datasets of different scales under real-world workloads. Compared to the five state-of-the-art approaches, \methodname{} achieves significant cost savings when training FedGCN models with thousands of FL participants. On average, it achieves comparable or up to 3.23\% higher test accuracy, while incurring 91.77\% and 85.59\% lower communication and lower computation cost, respectively. In this way, \methodname{} achieves significantly more advantageous trade-offs between efficiency and accuracy compared to existing approaches. 




\section{Related Work}

\subsection{FedGCN Training} 
Existing work on efficient intra-graph FedGCN training on large graphs can be divided into three branches. 
The first branch uses missing neighbor generation to obtain accurate node embeddings  \cite{zhang2021subgraph, zhang2021fastgnn}. However, it only focuses on improving prediction accuracy without considering the  computation and communication overhead caused by additional training of the generative model.
The second branch focuses on graph sampling approaches \cite{zhang2021subgraph}, \emph{e.g.}, FedGraph \cite{chen2021fedgraph} uses deep reinforcement learning (DRL) to select neighbor nodes for embedding aggregation. 
However, since it ignores the graph topology and heterogeneity of clients, it results in inaccurate node embeddings and large variance in the gradients.
Besides, it would incur large computation overhead as each local client needs to separately train two additional DRL networks.
The third branch periodically transfer cross-client neighbor node embeddings to reduce communication costs \cite{chen2021fedgraph, du2022federated, zhang2022fedego, deng2023graphfed}. 
However, they fail to capture the dynamics of model training to determine the optimal transfer period, resulting in inferior model performance.

\subsection{Sampling-based GCN Training} 
\begin{figure}[t!]
		\centering
		\subfigure[Coauthor]{\label{fig:testAcc_bytes_illusCoauthor_CS}
	\includegraphics[width = 0.45\linewidth, height=1.18in, trim=4 4 4 4]{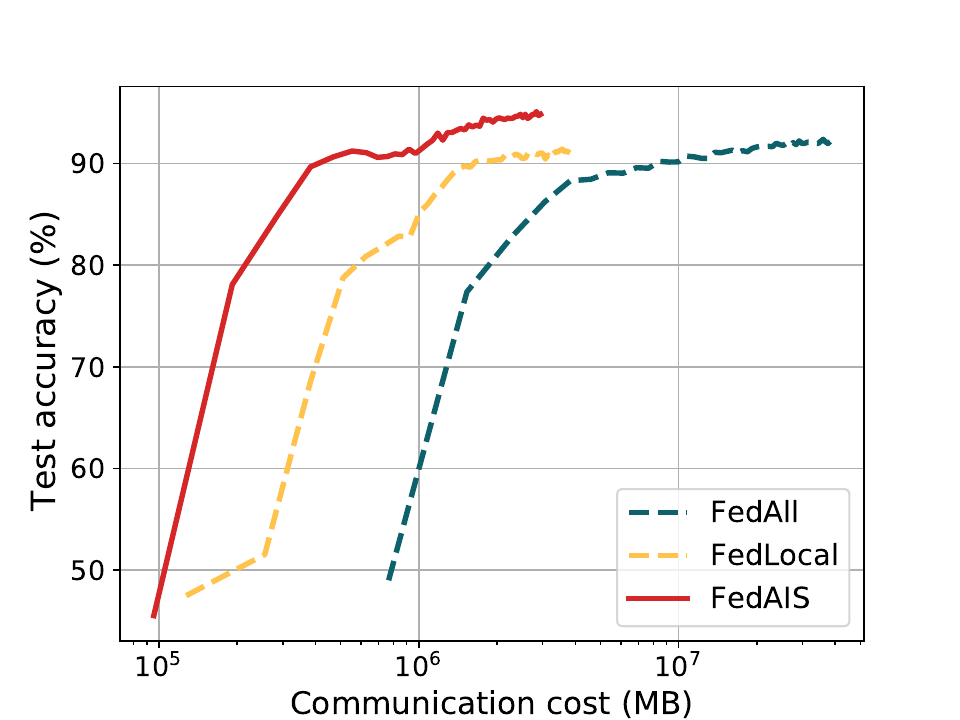}}
	\quad
	\subfigure[Reddit] {\label{fig:testAcc_bytes_illusReddit}
	\includegraphics[width = 0.45\linewidth, height=1.18in, trim=4 4 4 4] {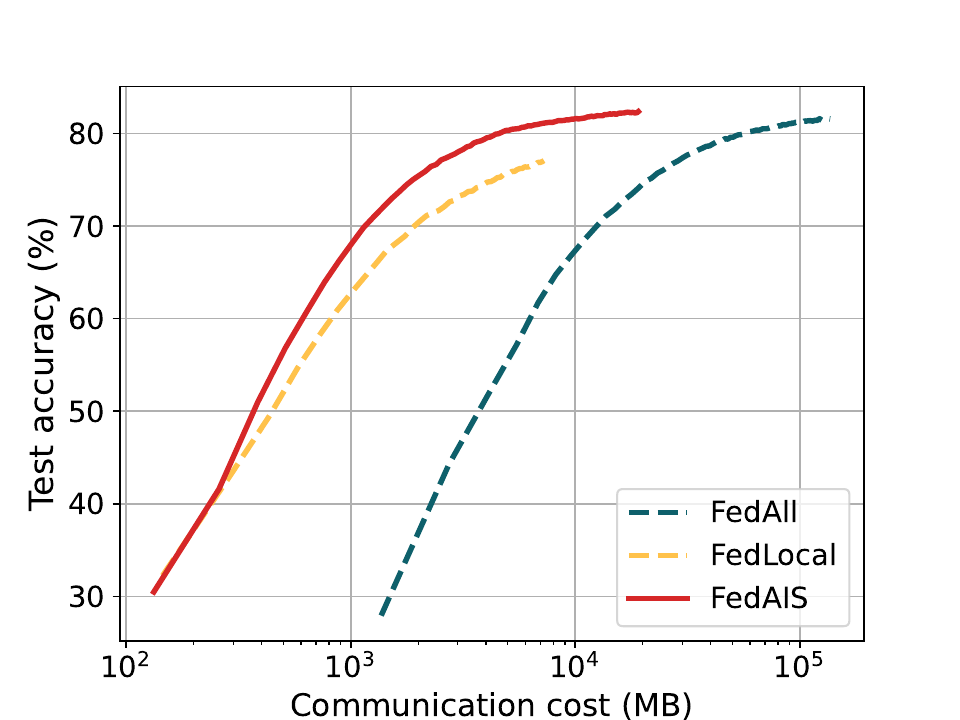}}
    	\vspace{-0.08in}
    	\caption{Test accuracy vs. communication costs. 
	}
	\label{fig:testAcc_bytes_illus}
\vspace{-0.1in} 
\end{figure}

One approach to scaling up GCN training is graph sampling, which can be categorized into: 1) node-wise sampling, 2) layer-wise sampling, and 3) subgraph sampling. 
Node-wise sampling methods 
\cite{hamilton2017inductive, cong2020minimal} iteratively sample a number of neighbors for each node based on specific probabilities (\emph{e.g.}, calculated based on node centrality).
Layer-wise sampling methods  
\cite{chen2018fastgcn, zou2019layer} independently sample a number of nodes for each GCN layer. Since multiple nodes are jointly sampled in each layer, the time cost of the sampling process is reduced by avoiding the exponential extension of neighbors. However, since nodes of different layers are sampled independently, some sampled nodes may have no connections with the ones in the previous layer, which would deteriorate the training performance. 
Subgraph sampling methods \cite{chiang2019cluster, zeng2019graphsaint} sample a number of subgraphs for each batch in GCN training. 
However, graph partitioning of large graphs is time-consuming and the model performance is highly sensitive to the cluster size. 
Those three categories of methods, however, require direct access to data features, which would risk privacy leakage of local clients in FL settings. 
Besides, those methods neglect communication costs and would incur substantial communication costs when the volume of the graph data is large. Thus, we propose \methodname{} to improve trade-offs between accuracy and efficiency (Fig. \ref{fig:testAcc_bytes_illus}). Here, FedLocal is the federated GraghSage \cite{hamilton2017inductive}, which conducts random selection of within-client neighbor nodes, where the cross-client neighbor information is ignored. FedPNS performs periodic selection of both within-client and cross-client neighbor nodes. 

\section{Problem Formulation}
There are two types of entities involved: a server $S$, and $K$ clients $\{1, 2, \cdots, K\}$. Each client $k$ owns a graph dataset $D_k=(G_k, Y_k)$, where $G_k=(V_k, E_k)$ is an undirected graph with  $n_k=|V_k|$ vertices, $|E_k|$ edges. $N=\sum_{k=1}^K n_k$ is the total number of all clients' local data samples.  
We focus on the task of node classification, where each vertex $v\in V_k$ is associated with a feature vector $x_v\in X_k$ and a label $y_v\in Y_k$. 
Given a $L$-layer FedGCN, let $f(h_v^{(L)}, \theta, y_{v})$, $F_k(h^{(L)}, \theta)$ denote loss functions of an individual sample $x_{v}$ and all samples on client $k$'s local model. 
$F(h^{(L)}, \theta)$ denotes the loss function of the global model. We formulate FedGCN learning as a distributed optimization problem:  
\begin{equation}
\label{eq:goal-FedGCN}
\small
\begin{split} 
    \theta^*=&\arg \min \{F(h^{(L)}, \theta)= \sum_{k=1}^{K}\frac{n_k}{N}F_k(h^{(L)}, \theta)\}, \\ 
    &{\rm where} 
	\, F_k(h^{(L)}, \theta)=\frac{1}{n_k}\sum_{v\in V_k} f(h_{v}^{(L)}, \theta, y_{v}),
\end{split} 
\end{equation} 
where the $l+1$-th graph convolution layer embedding $h_v^{(l+1)}$ of node $v\in V_k$ is defined as: 
\begin{equation}
\label{eq:separate}
\small 
    \begin{split}
        h_{v}^{(l+1)}=\sigma^{(l+1)}\Big(h_{v}^{(l)}, \underbrace{\{h_{w}^{(l)}\}_{w\in N(v)\cap V_k}}_{\text{Within-client nodes}} \cup \underbrace{\{h_{w}^{(l)}\}_{w\in N(v)\setminus V_k}}_{\text{Cross-client nodes}}\Big).
    \end{split}
\end{equation}
Here, $\sigma(\cdot)$ is the activation function. $N(v)$ denotes the set of neighbor nodes of $v$ and $h_v^{(1)}=x_v$. 
Suppose the average degree in a local graph $G_k$ is $d_k$. To calculate $h_{v}^{(l)}$ of node $v\in V_k$ in an $L$-layer FedGCN, on average the number of neighbors involved is $d_k^L$ \cite{chen2017stochastic}, which results in an exponential increase in computation and communication overhead with respect to $L$. 
Thus, local clients cannot afford to calculate all embedding terms $h_{v}^{(l)}$ which need to be  computed and transmitted recursively. 
Aggregating only within-client neighbor embeddings while ignoring cross-client information leads to inferior model performance \cite{chen2021fedgraph}. 




\section{Our FedAIS Approach}

\subsection{Joint Analysis of Variance and Overhead} 
\label{sec:error-ana}
To construct a global FedGCN model with fast convergence speed and low prediction error, we need to first analyze the sources of variances and biases when applying graph sampling strategies. Existing graph sampling approaches suffer from high variances and biases introduced to the stochastic gradients due to the approximation of node embeddings at different layers \cite{cong2020minimal, du2022federated}. 
We denote $\tilde{h}_v^{(l)}$ as the embedding approximation of node $v\in V_k$ in the $l$-th layer. 
Specifically, the variance of stochastic gradient estimator $\tilde{g}=\sum_{k=1}^K \frac{1}{|B_k|}\sum_{v\in B_k} \nabla f(\tilde{h}_v^{(L)}, \theta_t, y_v) $, can be decomposed as: 
\begin{equation}
\label{eq:variance-decomp}
    \mathbb{E}[||\tilde{g}-\nabla F(h^{(L)}, \theta) ||]=\mathbb{E}[||\tilde{g}-g ||]+\mathbb{E}[||g-\nabla{F}(h^{(L)}, \theta)||]. 
\end{equation}
$\mathbb{E}[||\tilde{g}-g ||]$ denotes the variance of estimated gradients to their exact values $g=\sum_{k=1}^K \frac{1}{n_k}\sum_{v\in V_k} \nabla f(h_v^{(L)}, \theta_t, y_v)$, resulting from the inner layer embedding approximation in forward propagation. The term $\mathbb{E}[||g-\nabla{F}(h^{(L)}, \theta)||]$ denotes the variance of mini-batch gradients to the full gradients due to the mini-batch sampling. 
\begin{assumption}
\label{ass:Lipschitz1}
    Let $F(h^{(L)}, \theta)$ be differentiable and $\lambda$-Lipschitz smooth and the value of $F(h^{(L)}, \theta)$ be bounded by a scalar $F_{inf}$. 
\end{assumption}

\begin{theorem}
\label{theorem:final-bound}
    Under Assumption \ref{ass:Lipschitz1}, if for all $v\in V_k$ and all $l\in \{1,2, \cdots, L-1\}$, the final output error of layer $L$ in training round $t\in [T]$ is bounded by:
\begin{equation} 
    ||\tilde{h}_{v}^{(L)}-h_{v}^{(L)}||\leq \sum_{l=1}^{L-1}\alpha_1^{L-l}\alpha_2^{L-l}|N(v)|^{L-l}. 
\end{equation}
\end{theorem} 
Theorem \ref{theorem:final-bound} lets us immediately derive an upper error bound for the estimated gradients, \emph{i.e.}, 
\begin{equation}
\label{eq:bound-gradient}
    \mathbb{E}[||\tilde{g}-g||]\leq \lambda ||\tilde{h}_v^{(L)}-h_v^{(L)}||. 
\end{equation} 

From the decomposition of variance in Eq. \eqref{eq:variance-decomp} and the upper error bound for the estimated gradients in Eq. \eqref{eq:bound-gradient}, we conclude that any graph sampling method introduces two sources of variance (\emph{i.e.}, the embedding approximation variance and the stochastic gradient variance). 
Therefore, to accelerate model convergence and reduce computation and communication overhead, both kinds of variance needs to be accounted in designing a graph sampling strategy. 

\subsection{System Overview}
\begin{figure}[!t]
	\begin{center}
		\includegraphics[width=0.98\linewidth,clip]{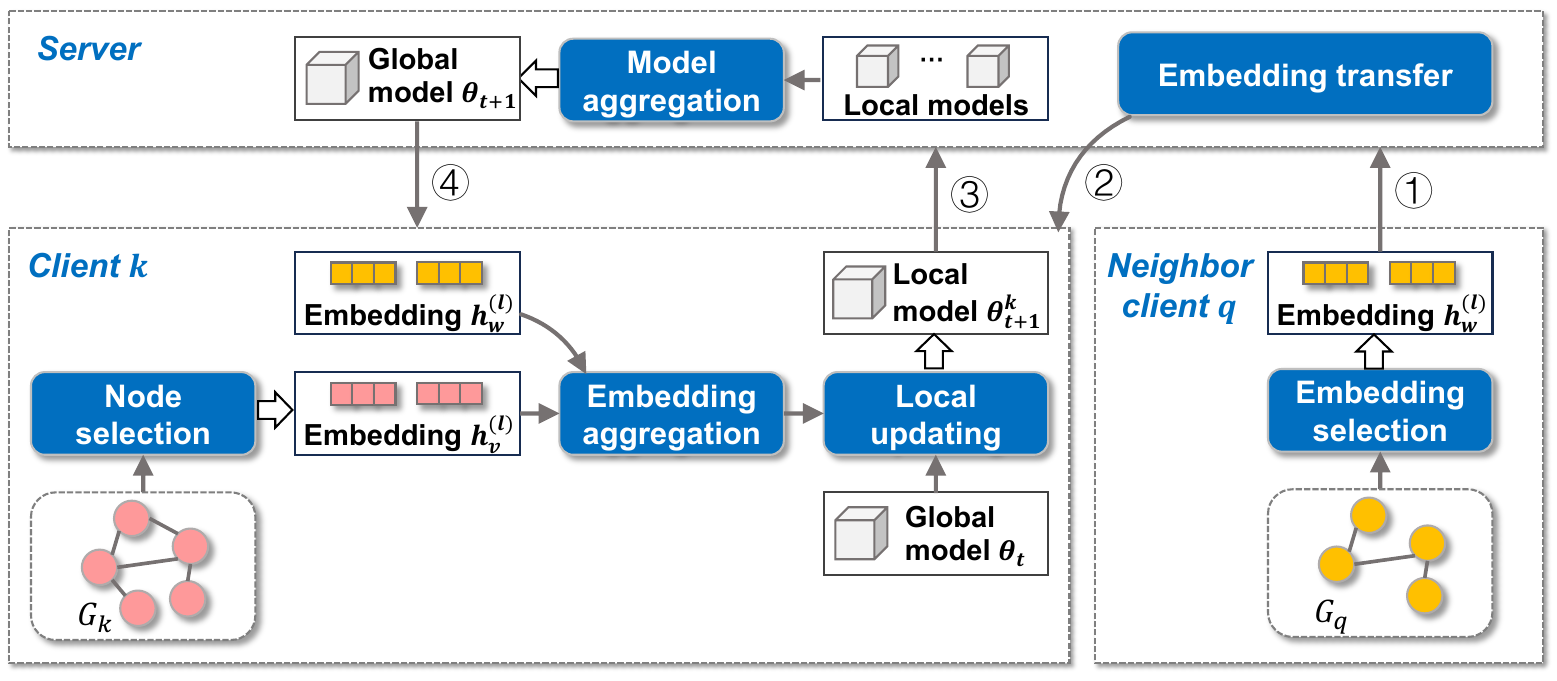}
  \vspace{-0.1in}
 		\caption{System overview of \methodname{}.  $\textcircled{1}$-$\textcircled{2}$ cross-client neighbor embeddings,     $\textcircled{3}$ local model, $\textcircled{4}$ global model $\theta_{t+1}$. }
		\vspace{-0.16in}
		\label{fig:system_overview}
	\end{center}
\end{figure}

\methodname{} consists of two modules (as shown in Fig. \ref{fig:system_overview}).
\begin{enumerate}
     \item \textbf{Historical Embedding-based Graph Sampling.} 
     In training round $t$,  client $k$ updates the importance scores of its local training samples based on historical embeddings and training losses of individual samples. Then, client $k$ selects its most influential samples to be used for FedGCN model training. Using historical embeddings and training losses helps reduce both embedding approximation variance and stochastic gradient variance, thereby enabling accurate FedGCN model training. 

     \item \textbf{Adaptive Embedding Synchronization and Model Updating.} 
    With the influence estimation and sampling results, each client $k$ updates its historical embeddings and performs node aggregation. Then, it updates its local model and estimates the next optimal synchronization interval via the joint analysis of the overhead and error-convergence. Finally, client $k$ sends the updated local model to the server, which then aggregates the local  models to produce the global model. 
     
 \end{enumerate}
\subsection{Historical Embedding-based Graph Sampling} 
While evaluating embedding $h_v^{(l)}$, it is prohibitively costly to calculate all $h_v^{(l)}$ terms since they need to be computed and transmitted recursively (\emph{i.e.}, we again need the embeddings $h_w^{(l-1)}$ of all $v$'s neighbor nodes $w$). 
To reduce computation and communication costs, we introduce the historical embedding for FedGCN as an affordable approximation. 
\begin{equation} 
\label{eq:history-calculate}
\small
    \begin{split}
        \tilde{h}_{v}^{(l+1)}= \sigma^{(l+1)}\Big(h_{v}^{(l)}, \{h_{w}^{(l)}\}_{w\in N(v)\cap B}\cup 
        \{\bar{h}_{w}^{(l)}\}_{w\in N(v)\setminus B} \Big),\\
        \{\bar{h}_w^{(l)}\}_{w\in N(v)\setminus B}= \underbrace{\{\bar{h}_{w}^{(l)}\}_{w\in N(v)\setminus B \cap V_k}\cup 
        \{\bar{h}_{w}^{(l)}\}_{w\in N(v)\setminus V_k}}_{\text{Historical embeddings}}.  
    \end{split}
\end{equation} 

Here, we separate the within-client neighbors into two parts: 1) within-client in-batch nodes $w\in N(v)\cap B$, which are part of the current batch $B_k\in V_k$; and 2) within-client out-of-batch nodes $w\in N(v)\setminus B_k \cap V_k$, which are part of the client $k$ but not included in the current batch $B_k$.  
For both neighbor nodes, we approximate their embeddings $h_w^{(l)}$ via historical embeddings $\bar{h}_w^{(l)}$ acquired in previous iterations.  Compared to the previous approach which incurs exponentially computation and communication costs, that incurred by historical embedding estimator increases linear with $L$, \emph{i.e.}, $O(\sum_{k=1}^K V_k | \cup_{v\in V_k} N(v)\cup \{v\}| \cdot LTJ)$ computation operations and $O(\sum_{k=1}^K \sum_{v\in V_k}\sum_{w\in N_{v}\setminus V_k}\sum_{l=1}^L d^l \cdot TJ)$ communication cost, where $T$ and $J$ are numbers of global training rounds and local training epochs.

Based on the historical embedding estimator and the variance analysis, we select training nodes that contribute most to the objective function and accelerate model convergence. It can be casted into the   following optimization problem, 
\begin{equation}
    \min_{P} \frac{1}{K}\sum_{k\in [K]}\frac{1}{n_k}\sum_{v\in V_k} \frac{||\nabla f(\tilde{h}_v^{(L)},\theta_t, y_v)||^2}{p_v}. 
\end{equation}
The most straightforward solution is to use the $L_2$ norm of gradient as the probability. However, it requires the calculation of $n_k$ derivatives for each client $k$ at each local epoch $j\in [\tau_t]$, which is computationally prohibitive \cite{li2021sample}. 
To solve this issue, we instead use the difference $\Delta_j= f(\tilde{h}_v^{(L)},\theta_{j+1}, y_v)-f(\tilde{h}_v^{(L)},\theta_j, y_v)$ of training losses between two consecutive local model updates to approximate the gradient $\Delta_j \approx \nabla f(\tilde{h}_v^{(L)},\theta_j, y_v)$. Then, client $k$ calculates the probability $p_v$ by normalizing the differences across its all training samples, 
\begin{equation}
\label{eq:prob-update1}
    p_{v}=\frac{||\Delta_j||}{\sum_{v\in V_k} ||\Delta_j||}, v\in V_k, j\in [\tau_t]. 
\end{equation}
Thus, the computational complexity is $O(n_k)$ since client $k$ only requires one forward propagation.

\subsection{Adaptive Embedding Synchronization}
\label{sec:fed-adap}

To analyze the effect of $\tau$ on the expected runtime, we consider the following delay model. In round $t$, the time taken by client $k$ to conduct a local model update at the $j$-th epoch is modeled as a random variable $c_{k,j}^t$ and the total communication delay is a random variable $o_{\tau}^t$. 
Then, the communication cost is $o_{k, \tau}^t b_t$, where $b_t$ is the average network bandwidth during round $t$. 
For the full synchronization, the total time to complete each round is $c_{\rm syn}=\max\{c_{1,1}^t, \cdots, c_{K,1}^t\}+o_{\tau}^t$, while for the periodic  synchronization, the average time to complete each round is $c_{\rm avg}=\max\{\bar{c}_1^t,\cdots,\bar{c}_K^t\}+o_{\tau}^t/\tau_t$ where $\bar{c}_k^t=\frac{1}{\tau_t}\sum_{j=1}^{\tau_t} c_{k,j}^t$. 
Consider the simplest case where $c_{k,\tau}^t=c$ and $o_{\tau}^t=o$ are constants, and $c/o$ is the ratio of communication delay to computation cost, which depends on the size of FedGCN model, network bandwidth and client computing capacity, \emph{etc}. 

\begin{assumption}
\label{ass:bound-variance}
    The stochastic gradient evaluated on the mini-batch $B_k$ with bounded variance, $\mathbb{E}[||\tilde{g}-g||]\leq \zeta^2$, where $\tilde{g}=\sum_{k=1}^K \frac{1}{|B_k|}\sum_{v\in B_k} \nabla f(\tilde{h}_v^{(L)}, \theta_t, y_v) $. 
\end{assumption}
\begin{theorem}
\label{theorem:syn-bound}
For periodic synchronization, under Assumption \ref{ass:Lipschitz1}-\ref{ass:bound-variance}, Theorem \ref{theorem:final-bound} and Eq. \eqref{eq:bound-gradient}, if the learning rate $\eta$ satisfies $\eta \lambda +\eta^2 \lambda^2\tau(\tau-1)\leq 1$, 
and $\theta_0$ is the initial model generated by the server,
then after total $c_{\rm total}$ runtime, the minimal expected squared gradient norm is bounded by 
\begin{equation} 
\label{eq:error-bound}
    \frac{2 (F(\tilde{h}^{(L)},\theta_0)-F_{inf})}{\eta c_{\rm total}}(c+\frac{o}{\tau}) +\eta^2\lambda^2\zeta^2(\tau-1).
\end{equation} 
\end{theorem} 
From the optimization error bound in Eq. \eqref{eq:error-bound}, the error-runtime trade-off for different synchronization communication intervals can be derived. While a larger $\tau$ reduces the runtime per iteration and makes the first term in Eq. \eqref{eq:error-bound} smaller, it also adds noise and increases the last term. 

Then, we determine the optimal embedding synchronization interval $\tau$ to minimize the optimization error for each training batch. 
We start with infrequent cross-client embedding synchronization for improved convergence speed, and gradually transiting to higher embedding synchronization frequencies to reduce the prediction error of the learned global model. 
Specifically, at each training round $t$, the server selects the optimal  embedding transmission interval that achieves the fast test loss decay of the global model $\theta_t$ for the next interval. 
Theorem \ref{theorem:syn-bound} illustrates that there is an optimal value $\tau_t$ that minimizes the optimization error bound at round $t$ between the server and all selected clients,  
\begin{equation}
    \label{eq:tau-cal}\tau_t=\sqrt{\frac{2(F(\tilde{h}^{(L)},\theta_{t})-F_{inf})o}{\eta^3 c_{\rm total} \lambda^2 \zeta^2 }}. 
\end{equation}
It can be observed from Eq. \eqref{eq:tau-cal} that the generated synchronization period sequence decreases along with the objective value on the test set when the learning rate is fixed. It is consistent with the intuition that the trade-off between error-convergence and  overhead varies over time. Compared to the initial training phase, the benefit of using a large synchronization interval diminishes as the model converge since a lower error is preferred in the latter training phase. 
In some scenarios where the Lipschitz constant $\lambda$ and the gradient variance bound $\zeta^2$ are unknown and estimating these constants are difficult due to the highly non-convex and high-dimensional loss surface. As an alternative, we propose a simpler rule where we approximate $F_{inf}$ by 0, and divide Eq. \eqref{eq:tau-cal} by $\tau_0$ to obtain the basic synchronization interval,
\begin{equation}
\label{eq:interval-update}
    \tau_t=\Big\lceil \sqrt{\frac{F(\tilde{h}^{(L)},\theta_{t})}{{F(\tilde{h}^{(L)},\theta_0)}}}\tau_0\Big\rceil,  
\end{equation} 
where $\lceil r \rceil$ is the ceil function to round $r$ to the nearest integer. In practical implementations, we take the test loss as the objective function value and the average batch number $\sum_{k=1}^K n_kB/K$ as the initial synchronization period $\tau_0$, both of which can be easily obtained during training.




\begin{algorithm}[!b]
 \SetAlgoVlined
\small{
    \caption{\methodname{}}
    \label{alg:FedAIS}
    \SetKwInOut{Input}{Input}\SetKwInOut{Output}{Output}
    \Input{Initial probability $p_{k,i}^0=\frac{1}{n_k}$, sampling ratio $r_k$ 
    }
    \Output{The optimal global model $\theta^*$}
    //At the FL Server:\\
     {Initialize global model $\theta_0$;\\
    \For{each round $t\in\{1, \cdots, T\}$}{
    $M_t\leftarrow$ randomly select $m$ clients;\\ 
    	 \For{each client $k\in M_t$}{
	    $\theta_{t+1}^k \leftarrow$ \texttt{LocalUpdate}($k, \theta_t, \tau_t$);\\ 
	    } 
	    $\theta_{t+1}=\frac{1}{m}\sum_{k\in M_t} \theta_{t+1}^k$; // update global model \\ 
     Calculate $\tau_{t+1}$ with Eq. \eqref{eq:interval-update}; // update interval 
		}
     }
     //At FL Client $k\in [K]$:\\
	 \SetKwFunction{Flmu}{LocalUpdate}
      \SetKwProg{Fn}{Function}{:}{}
      \Fn{\Flmu{$k,\theta_t, \tau_t$}}{
      Calculate loss difference $\Delta_{j-1}$;\\
    Update selection probability $p_{t}^v=\frac{||\Delta_{j-1}||}{\sum_{v\in V_k} ||\Delta_{j-1}||}$; \\
    \For{each local epoch $j\in \{1, 2, \cdots, J\}$}{
    Select a batch $B$ of $\frac{n_k}{|B|}r_k$ samples $x_{k,i}\propto p_{k,i}^t$;\\ 
    \If{$j\%\tau_t==0$}{
    Calculate $\tilde{h}_i^{(l+1)}$ with Eq. \eqref{eq:history-calculate};\\  
    Update historical embedding $\bar{h}_i^{(L)}$;\\ 
    }
    $\theta_{j}^k \leftarrow \theta_{j-1}^k - \eta \frac{1}{|B|}\sum_{v\in B} \nabla f(\tilde{h}_{v}^{(L)},y_{v})$;\\ 
    }  
    \textbf{Return} $\theta_J^k$;\\
        } 
        }
\end{algorithm}

\subsection{Implementation} 
The proposed \methodname{} is illustrated in Algorithm \ref{alg:FedAIS}. 
Specifically, in the $t$-th global round, the server randomly selects a set $M_t$ of $m$ clients, and distributes the current model $\theta_t$ to them (Lines 4-6). Each selected client $k$ calculates the loss for each sample $i\in V_k$ and updates its selection probability $p_{k,i}^t$ (Lines 11-12). 
Then, during each local epoch $j$, client $k$ selects a batch $B_k$ of samples with $x_i \propto p_{k,i}^t, i\in V_k$.
For each layer of FedGCN, when the number of local batch training epoch $j$ satisfies $j\% \tau_t==0$, client $k$ firstly calculates $\tilde{h}_i^{(l+1)}$ with Eq. \eqref{eq:history-calculate} by aggregating embeddings of both within-client neighbors and cross-client neighbors. Then, it performs embedding synchronization by asking neighbor client $q\in Q$ to update the selected cross-client neighbor embeddings and transmit them back (Lines 13-18). 
Then, client $k$ updates its local model $\theta_j^k$ and sends  $\theta_J^k$ to the server. 
The server aggregates updates by conducting model aggregation and updating interval $\tau_{t+1}$ (Lines 7-8). In this way, the server and clients collaboratively train a FedGCN model $\theta^*$ with high prediction accuracy and low overhead.

\subsection{Convergence Analysis}
Without loss of generality, we analyze an arbitrary interval sequence $\{\tau_1, \cdots, \tau_R\}$ with $R$ synchronization iterations. 
\begin{theorem} {(Convergence of FedAIS)} Suppose the learning rate $\eta$ remains the same as $R\rightarrow\infty$,  
    \begin{equation}
    \label{eq:learning-rate}
        \sum_{r=0}^R \eta_r \tau_r\rightarrow\infty, \sum_{r=0}^R \eta_r^2 \tau_r<\infty, \sum_{r=0}^R \eta_r^3 \tau_r^2<\infty,
    \end{equation}
    The global model $\theta$ is guaranteed to converge to: 
    \begin{equation}
        \mathbb{E}\Big[\frac{\sum_{r=0}^{R-1}\eta_r\sum_{k=1}^{\tau_r}||\nabla F(\tilde{h}^{(L)},  \theta_{\sum_{i=0}^{r-1}\tau_i+k})||}{\sum_{r=0}^{R-1}\eta_r \tau_r} \Big]\rightarrow 0. 
    \end{equation}
\end{theorem}
The key idea of proof is as follows. To understand the condition \eqref{eq:learning-rate}, we consider the case when $\tau_0=\cdots =\tau_R$ is a constant. Then, the converge condition is identical to that for mini-batch SGD: $\sum_{r=0}^R \eta_r \rightarrow \infty, \sum_{r=0}^R \eta_r^2 < \infty$. Provided that the sequence of communication periods is bounded, the learning rate in mini-batch SGD can be easily adjusted to satisfy condition \eqref{eq:learning-rate}. 
Specifically, when the communication period sequence decreases, the last two terms in \eqref{eq:learning-rate} become easier to satisfy, and the differences between the objective values of two consecutive rounds are bounded. The full proof of \methodname{} convergence is presented in Appendix.

\label{sec:sampling}


\section{Experiment Evaluation}
\subsection{Experimental Settings}
\label{sec:exp_configuration}
\begin{table}[!t]
\centering
\vspace{-0.1in}
\caption{Statistics of the datasets. $\triangle E$ denotes the total number of  cross-client edges.} 
\label{tab:dataset} 
\resizebox{1\linewidth}{!} {
\renewcommand{\arraystretch}{1.04}
\begin{tabular}{c|c|c|c|c|c} 
\toprule
\textbf{Dataset}& \textbf{Coauthor}  & \textbf{Pubmed} & \textbf{Yelp} & \textbf{Reddit} & \textbf{Amazon2M}\\ 
\hline 
$V$    &   18,333    &  	
19,717   &   716,847   & 232,965   & 2,449,029   \\
$E$   &  163,788     &  88,648 &13,954,819 &	
114,615,892 &  61,859,140 \\ 
\# features &   6,805     &    500 & 300     & 602 & 100\\
\# classes  &   15 &  3  & 100     & 41 & 47 \\ 
Train/ Val/ Test  &   0.8/ 0.1/ 0.1     &  0.8/ 0.1/ 0.1  & 0.75/ 0.10/ 0.15     & 0.66/ 0.10/ 0.24 & 0.8/ 0.1/ 0.1 \\ 
\hline
\multicolumn{6}{c}{\textbf{100 clients}} \\ \hline 
$V_k$    &   146  &  158 &  5,376   & 1,538  & 19,592 \\ 
$E_k$    &  173  & 879 &138,815 & 1,140,985 & 610,748 \\ 
$\triangle  E$    &    1,030 &   747  &73,230 &	
517,533   & 784,277  \\ 
\bottomrule 
\end{tabular} 
}
\vspace{-0.06in}
\end{table}
\begin{table*}[]
\centering 
\caption{Performance comparison for training different FedGCN models on various datasets. } 
\label{tab:perf-metric}
\resizebox{1.0\linewidth}{!}{
\renewcommand{\arraystretch}{1.0}
\begin{tabular}{|c|c|c|c|c|c|c|c|c|c|c|c|}
\toprule 
\textbf{Method} & \textbf{Metric} & \multicolumn{10}{|c|}{\textbf{Performance results (\%) $\pm$ standard deviations} }         \\ \hline 
&  & \multicolumn{2}{c|}{Coauthor} & \multicolumn{2}{c|}{Pubmed} & \multicolumn{2}{c|}{Yelp} & \multicolumn{2}{c|}{Reddit}& \multicolumn{2}{c|}{Amazon2M} \\ \cline{2-12} 
& & iid  & non-iid & iid  & non-iid     & iid  & non-iid  & iid & non-iid & iid & non-iid  \\ \hline  
\multirow{3}{*}{FedAll} & testAcc   & \textbf{89.98{\footnotesize$\pm$ 0.78}} &84.46{\footnotesize$\pm$ 1.08} & 87.74{\footnotesize$\pm$ 2.16}  &\textbf{86.42{\footnotesize$\pm$ 1.07}} &91.57{\footnotesize$\pm$ 1.29}& 89.69{\footnotesize$\pm$ 2.08} & 82.38{\footnotesize$\pm$ 0.61}& 81.66{\footnotesize$\pm$ 0.62} & 71.46{\footnotesize$\pm$ 1.03}&67.89{\footnotesize$\pm$ 0.82} \\ 
& F1-score & \textbf{78.38{\footnotesize$\pm$ 1.25}}  & \textbf{73.03{\footnotesize$\pm$ 1.13}}  &85.62{\footnotesize$\pm$ 0.87}  & 84.51{\footnotesize$\pm$ 0.91} &29.03{\footnotesize$\pm$ 1.14}  &27.71{\footnotesize$\pm$ 1.41} &76.34 {\footnotesize$\pm$ 1.12}&74.89{\footnotesize$\pm$ 1.15} &29.89{\footnotesize$\pm$ 0.85} & 25.82{\footnotesize$\pm$ 0.79} \\ 
& AUC  & 92.35{\footnotesize$\pm$ 1.32} & \textbf{75.26{\footnotesize$\pm$ 1.28}} & \textbf{96.14{\footnotesize$\pm$ 0.14}} &  96.03{\footnotesize$\pm$ 0.74} &  75.45{\footnotesize$\pm$ 2.03} & 73.36{\footnotesize$\pm$ 1.83} &\textbf{95.54{\footnotesize$\pm$ 0.05}}&92.41{\footnotesize$\pm$ 1.02} &83.62{\footnotesize$\pm$ 0.65} &60.12{\footnotesize$\pm$ 0.87}\\ 
\hline 
\multirow{3}{*}{FedRandom} & testAcc   & 87.09{\footnotesize$\pm$ 0.78} &81.53{\footnotesize$\pm$ 1.28} & 87.16{\footnotesize$\pm$ 2.16}  &84.92{\footnotesize$\pm$ 1.07} &92.25{\footnotesize$\pm$ 2.29}& 86.79{\footnotesize$\pm$ 1.12} & 79.34{\footnotesize$\pm$ 0.52}& 77.52{\footnotesize$\pm$ 0.91} & 71.78{\footnotesize$\pm$ 0.49} & 65.45{\footnotesize$\pm$ 0.64} \\ 
& F1-score & 72.95{\footnotesize$\pm$ 1.05}  & 68.13{\footnotesize$\pm$ 2.13}  &83.62{\footnotesize$\pm$ 0.87}  & 81.32{\footnotesize$\pm$ 0.91} &29.61{\footnotesize$\pm$ 2.14}  &27.21{\footnotesize$\pm$ 1.41} &72.34 {\footnotesize$\pm$ 1.52}&70.56{\footnotesize$\pm$ 1.05}& 30.54{\footnotesize$\pm$ 0.68} &24.43{\footnotesize$\pm$ 0.48}  \\ 
& AUC  & 88.58{\footnotesize$\pm$ 1.27} & 70.76{\footnotesize$\pm$ 0.82} & 96.12{\footnotesize$\pm$ 0.16} &  92.06{\footnotesize$\pm$ 0.74} &  74.84{\footnotesize$\pm$ 2.03} & 71.58{\footnotesize$\pm$ 0.79} &93.66{\footnotesize$\pm$ 0.05}&88.42{\footnotesize$\pm$ 1.08}& 83.16{\footnotesize$\pm$ 0.54} & 56.98{\footnotesize$\pm$ 1.12}\\ 
\hline 
\multirow{3}{*}{FedSage+} & testAcc   & 85.02{\footnotesize$\pm$ 0.78} &81.95{\footnotesize$\pm$ 1.03} & 86.82{\footnotesize$\pm$ 1.16}  &85.08{\footnotesize$\pm$ 1.14} &83.45{\footnotesize$\pm$ 0.79}& 82.79{\footnotesize$\pm$ 0.43} & 78.34{\footnotesize$\pm$ 0.28}& 74.52{\footnotesize$\pm$ 0.62} & 71.47{\footnotesize$\pm$ 0.79} & 66.42{\footnotesize$\pm$ 0.81} \\ 
& F1-score & 76.78{\footnotesize$\pm$ 1.05}  & 54.93{\footnotesize$\pm$ 0.63}  &84.62{\footnotesize$\pm$ 0.87}  & 83.21{\footnotesize$\pm$ 0.91} &30.12{\footnotesize$\pm$ 0.74}  &29.84{\footnotesize$\pm$ 0.81} &\textbf{76.54{\footnotesize$\pm$ 0.97}}&66.56{\footnotesize$\pm$ 1.05}&29.89{\footnotesize$\pm$ 0.38}&27.38{\footnotesize$\pm$ 0.37}  \\ 
& AUC  & 89.29{\footnotesize$\pm$ 1.17} & 74.45{\footnotesize$\pm$ 0.92} & 90.83{\footnotesize$\pm$ 0.58} &  89.06{\footnotesize$\pm$ 0.49} &  71.26{\footnotesize$\pm$ 1.03} & 72.27{\footnotesize$\pm$ 0.72} &94.76{\footnotesize$\pm$ 0.55}&92.42{\footnotesize$\pm$ 1.18}&83.62{\footnotesize$\pm$ 0.48}&58.24{\footnotesize$\pm$ 0.42} \\ 
\hline
    \multirow{3}{*}{FedPNS } & testAcc  &87.03{\footnotesize$\pm$ 0.21} & 81.65{\footnotesize$\pm$ 0.93} &87.38{\footnotesize$\pm$ 1.23}  & 85.87{\footnotesize$\pm$ 1.05} &90.92{\footnotesize$\pm$ 0.59} &85.97{\footnotesize$\pm$ 0.82} &82.25{\footnotesize$\pm$ 1.26}&80.24{\footnotesize$\pm$ 0.73}& 71.36{\footnotesize$\pm$ 0.78}& 66.74{\footnotesize$\pm$ 0.43} \\ 
& F1-score  & 72.58{\footnotesize$\pm$ 2.34} &68.32{\footnotesize$\pm$ 1.62} &85.96{\footnotesize$\pm$ 2.74} &\textbf{84.81{\footnotesize$\pm$ 2.46}} &28.99{\footnotesize$\pm$ 1.02}& 28.01{\footnotesize$\pm$ 1.24} &76.10{\footnotesize$\pm$ 0.62}& 73.63{\footnotesize$\pm$ 0.62}& 29.63{\footnotesize$\pm$ 0.78}& 25.95{\footnotesize$\pm$ 1.07}  \\ 
& AUC  & 87.69{\footnotesize$\pm$ 2.37}& 72.81{\footnotesize$\pm$ 1.72}&95.87{\footnotesize$\pm$ 1.29} & 93.86{\footnotesize$\pm$ 1.09} &75.22{\footnotesize$\pm$ 0.25} &73.74{\footnotesize$\pm$ 1.02} 
 &95.07{\footnotesize$\pm$ 1.28} &91.12{\footnotesize$\pm$ 0.36}& 83.39{\footnotesize$\pm$ 0.52}& 58.19{\footnotesize$\pm$ 0.62} \\ 
\hline 
\multirow{3}{*}{FedGraph}  & testAcc & 88.09{\footnotesize$\pm$ 1.06}  & 83.18{\footnotesize$\pm$ 1.25} &  87.82{\footnotesize$\pm$ 0.97}& 85.17{\footnotesize$\pm$ 0.84} & 90.98{\footnotesize$\pm$ 0.82} &87.41{\footnotesize$\pm$ 0.58} &82.18{\footnotesize$\pm$ 0.75} &80.12{\footnotesize$\pm$ 1.14}& 71.75{\footnotesize$\pm$ 0.62} & 66.93{\footnotesize$\pm$ 0.47} \\  
& F1-score  & 73.19{\footnotesize$\pm$ 1.24} &68.03{\footnotesize$\pm$ 1.86} &85.69{\footnotesize$\pm$ 1.04} & 83.58{\footnotesize$\pm$ 1.41} &31.39{\footnotesize$\pm$ 1.22} & 29.16{\footnotesize$\pm$ 1.02} &75.88{\footnotesize$\pm$ 0.83}& 71.42{\footnotesize$\pm$ 1.04}& 30.21{\footnotesize$\pm$ 0.82} & 25.43{\footnotesize$\pm$ 0.93}  \\  
& AUC  &  88.85{\footnotesize$\pm$ 0.78}  & 72.15{\footnotesize$\pm$ 1.43}  &95.89{\footnotesize$\pm$ 1.62} & 93.97{\footnotesize$\pm$ 1.27} &77.75{\footnotesize$\pm$ 0.51} 
 &73.31{\footnotesize$\pm$ 0.89}  & 94.96{\footnotesize$\pm$ 1.17}&92.61{\footnotesize$\pm$ 0.72}& 82.82{\footnotesize$\pm$ 0.78}& 59.89{\footnotesize$\pm$ 0.53}  \\ 
\hline 
\multirow{3}{*}{\methodname{}} & testAcc &88.12{\footnotesize$\pm$ 0.12}  & \textbf{85.49{\footnotesize$\pm$ 0.79}}  & \textbf{88.36{\footnotesize$\pm$ 0.59}} & 85.26{\footnotesize$\pm$ 1.07} &\textbf{94.12{\footnotesize$\pm$ 0.17}}  &\textbf{91.13{\footnotesize$\pm$ 0.72}} &\textbf{82.48{\footnotesize$\pm$ 0.18}}  &\textbf{81.84{\footnotesize$\pm$ 0.93}}& \textbf{71.84{\footnotesize$\pm$ 0.31}}& \textbf{67.99{\footnotesize$\pm$ 0.58}} \\  
& F1-score  & 74.86{\footnotesize$\pm$ 1.07} & 69.16{\footnotesize$\pm$ 0.37}  & \textbf{86.34{\footnotesize$\pm$ 0.21}}    & 83.72{\footnotesize$\pm$ 0.83}  &\textbf{31.65{\footnotesize$\pm$ 0.26}} &\textbf{30.87{\footnotesize$\pm$ 0.69}} &76.16{\footnotesize$\pm$ 0.23} &\textbf{74.69{\footnotesize$\pm$ 0.53}}& \textbf{30.84{\footnotesize$\pm$ 0.52}}& \textbf{27.52{\footnotesize$\pm$ 0.81}} \\ 
& AUC  & \textbf{92.52{\footnotesize$\pm$ 0.12}} & 73.75{\footnotesize$\pm$ 1.04} & 95.97{\footnotesize$\pm$ 0.31} &\textbf{96.28{\footnotesize$\pm$ 0.94}} & \textbf{79.96{\footnotesize$\pm$ 0.12}}
&74.35{\footnotesize$\pm$ 0.72}&95.26{\footnotesize$\pm$ 1.05}&\textbf{92.47{\footnotesize$\pm$ 0.82}}& \textbf{84.16{\footnotesize$\pm$ 0.26}} & \textbf{60.72{\footnotesize$\pm$ 0.43}}\\
\bottomrule  
\end{tabular}
}
\vspace{-0.1in}
\end{table*}
\begin{figure*}[t!]
	\begin{center}
	    \begin{minipage}[t]{0.48\linewidth}
	        \subfigure[\texttt{FedReddit}]{\label{fig:comm_testAcc_bytesReddit}
	\includegraphics[width = 0.47\linewidth, height=1.18in, trim=4 4 4 4]{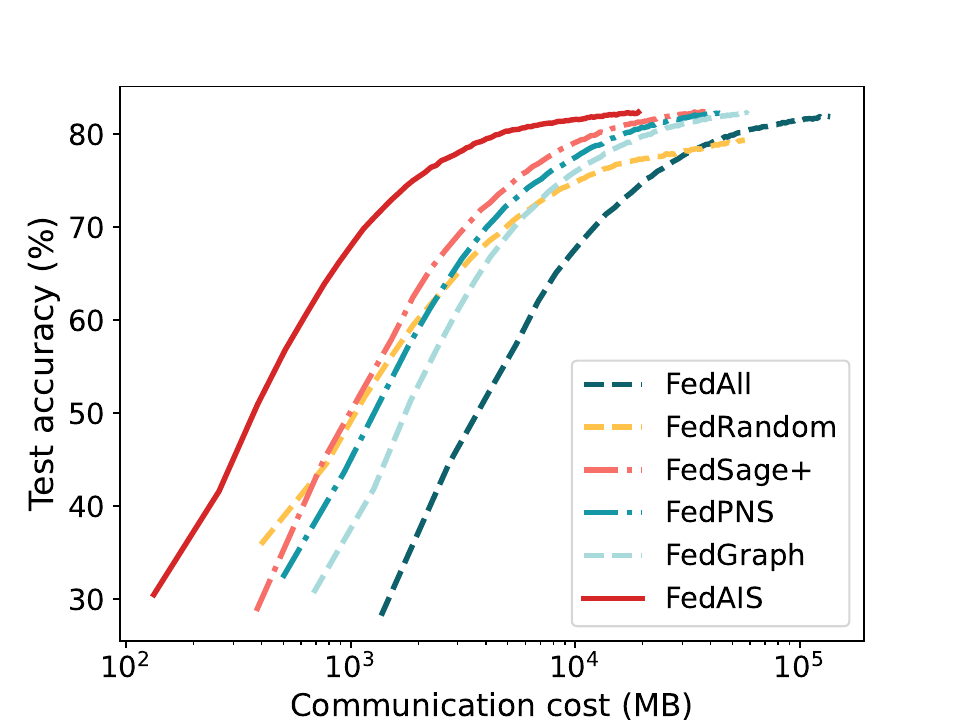}}
	\subfigure[\texttt{FedAmazon}] {\label{fig:comm_testAcc_bytesAmazon2M}
	\includegraphics[width = 0.47\linewidth, height=1.18in, trim=4 4 4 4]{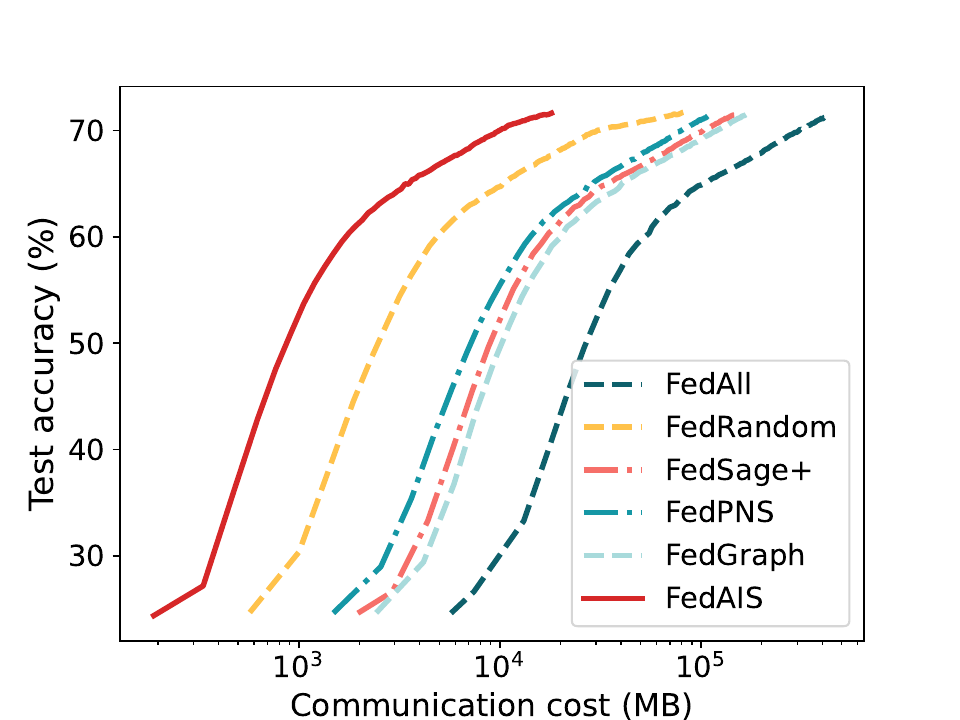}}
    	\vspace{-0.1in}
    	\caption{Accuracy scores with sizes of total communication cost for training different FedGCN models. 
	}
	\label{fig:testAcc_bytes}
	    \end{minipage}
     \quad 
        \begin{minipage}[t]{0.48\linewidth}
           \subfigure[Computation cost]{\label{fig:comp_cost}
	\includegraphics[width = 0.47\linewidth, height=1.18in, trim=4 4 4 4]{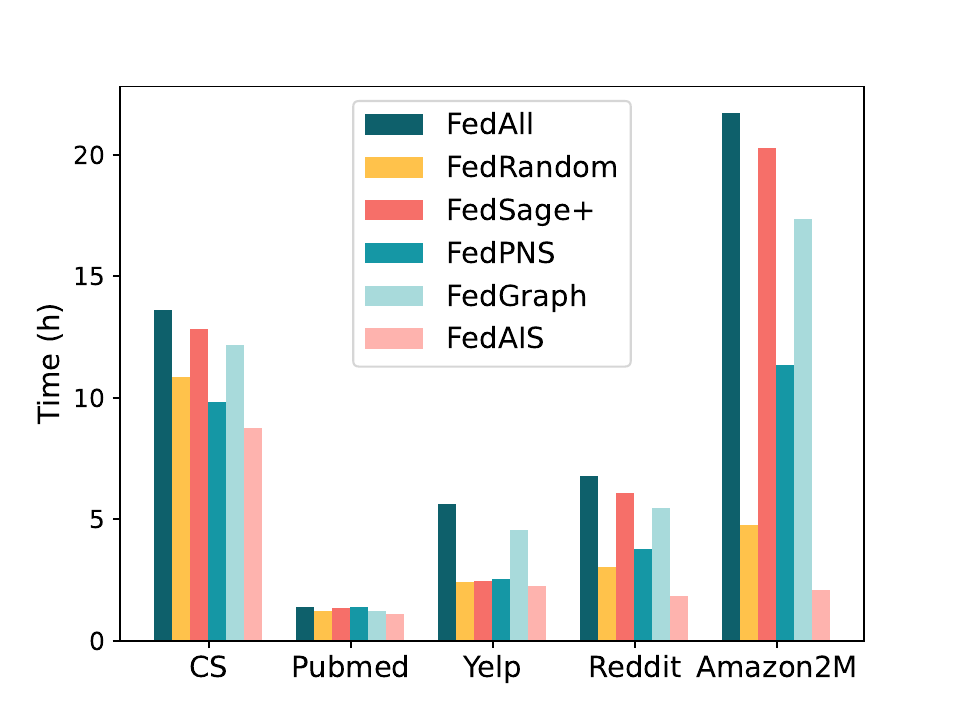}}
	\subfigure[Communication cost] {\label{fig:communication_cost}
	\includegraphics[width = 0.47\linewidth, height=1.18in, trim=4 4 4 4]{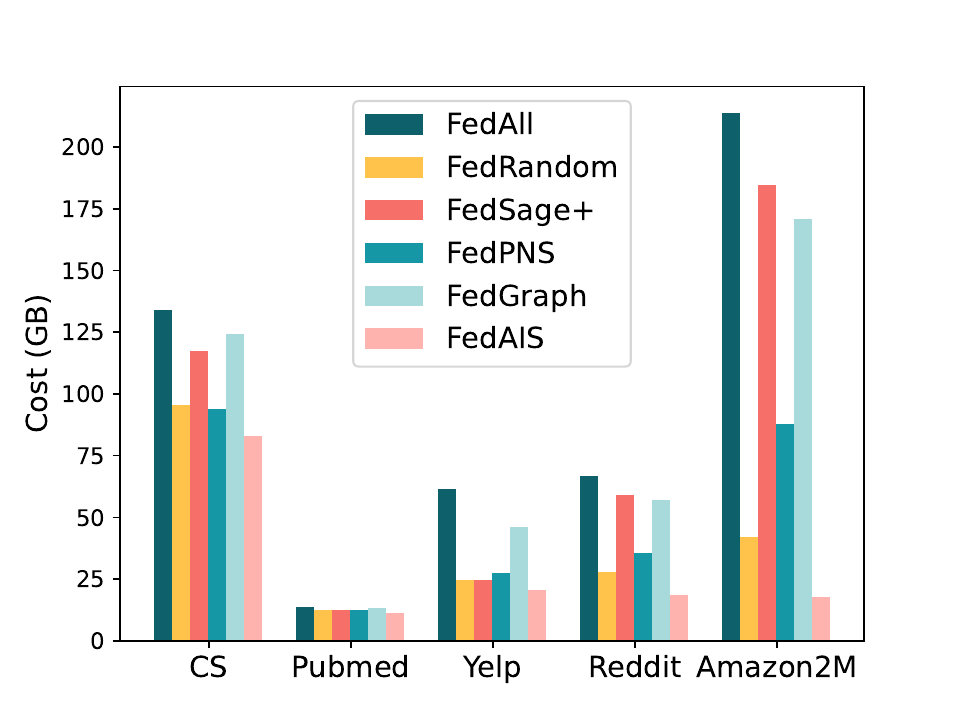}}
    	\vspace{-0.1in}
    	\caption{The total computation and communication costs for training various FedGCN models.
	}
	\label{fig:comp_comm_cost} 
        \end{minipage}
    \end{center}
\vspace{-0.14in} 
\end{figure*}
\textbf{Implementation.} We implemented \methodname{} and deployed it in an FL system consisting of one server and 100 clients. To further investigate the performance of \methodname{} in large-scale FL systems, we also tested it in an environment with up to 1,000 clients. Our implementation is based on Python 3.11 and Pytorch Geometric 2.0.1 \cite{Lenssen}. All the experiments are performed on Ubuntu 20.04 operating system equipped with a 32-core AMD Ryzen Threadripper PRO 5965WX @ 3.800GHz CPU, 192G of RAM and a NVIDIA RTX A5000 GPU with 24GB memory. 

\textbf{Datasets and Models.} 
We use five real-world graph datasets of different scales, \emph{i.e.},  
Coauthor \cite{shchur2018pitfalls}, 
Pubmed \cite{sen2008collective}, Yelp \cite{zeng2019graphsaint}, Reddit \cite{hamilton2017inductive}, Amazon2M \cite{hu2020open}. 
We partition training/ validation sets over 100 clients in both independent and identically distributed (iid) setting and non-iid setting.  
We use a non-iid partition by $p_i \sim Dir_k(\alpha)$, $\alpha=0.5$ with a Dirichlet distribution and allocate a $p_{i,k}$ ratio of instances of class $i$ to client $k$   \cite{li2022federated, yurochkin2019bayesian}. 
Since the original graph is extremely dense, we downsample the edges in local subgraphs by 50\% \cite{hamilton2017inductive}. 
The test dataset is located at the server and the statistics of the datasets are presented in Table \ref{tab:dataset}. 
We use the widely adopted GraphSage model \cite{hamilton2017inductive} with FedAvg to construct FedGCN models: \texttt{FedAuthor}, \texttt{FedPubmed}, \texttt{FedYelp}, \texttt{FedReddit},  \texttt{FedAmazon} \cite{hu2020open}. 
Each model has two hidden layers with 256, 128 neurons. 
We set the ratio of sample selection to 0.7, the number of neighbors sampled to 10, and the minimum embedding synchronization interval $\tau_0$ to 2 batch training epochs. 
We use Adam as the optimizer with the weight decay 0.001 and ReLU as the activation function. 
We set the learning rate $\eta=0.001$, the fixed batch number is 10, and the global warm-up round is 1. We conduct training until a pre-specified test accuracy is reached, or a maximum number of iterations has elapsed (\emph{e.g.}, 100 rounds). We perform 5-fold cross validation and report the average results. 

\textbf{Comparison Baselines.} 
\textbf{1) FedAll}: It conducts training using all local samples  and performs random neighbor node selection of both local subgraph neighbors and cross-client neighbors. 
\textbf{2) FedRandom}: It performs random selection for both local samples and neighbor nodes in each batch training.
\textbf{3) FedSage+} \cite{zhang2021subgraph}: It proposes a GNN-based neighbor generative model for each client to predict the features of each node's cross-client neighbors.
\textbf{4) FedPNS} \cite{du2022federated}: It conducts training using all local samples and conducts periodic neighbor node selection for cross-client neighbor nodes. We set the periodic interval to 2 local batch training epochs. 
\textbf{5) FedGraph} \cite{chen2021fedgraph}: It performs FL training using all local samples and selects local subgraph neighbors and cross-client neighbors by adjusting sampling policies based on DRL. 

\begin{figure*}[t!]
	    \begin{minipage}[t]{0.48\linewidth}
	        \subfigure[Test accuracy, \texttt{FedReddit}]{\label{fig:testAcc_ablation_bytesReddit}
	\includegraphics[width = 0.47\linewidth, height=1.18in, trim=4 4 4 4]{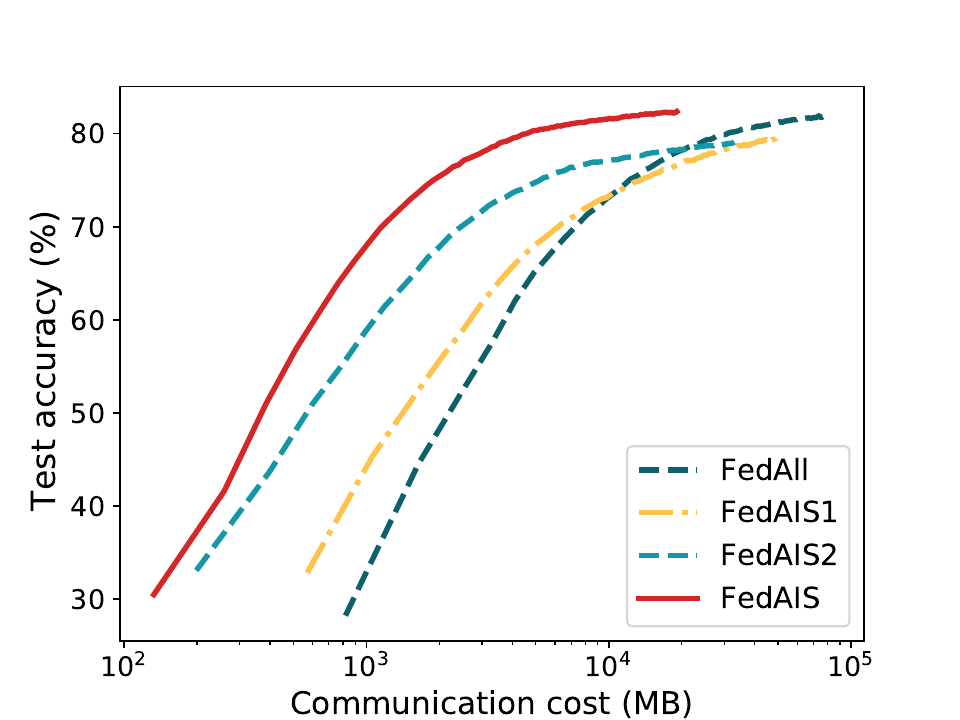}}
	\subfigure[Communication cost] {\label{fig:comm-ablation}
	\includegraphics[width = 0.47\linewidth, height=1.18in, trim=4 4 4 4]{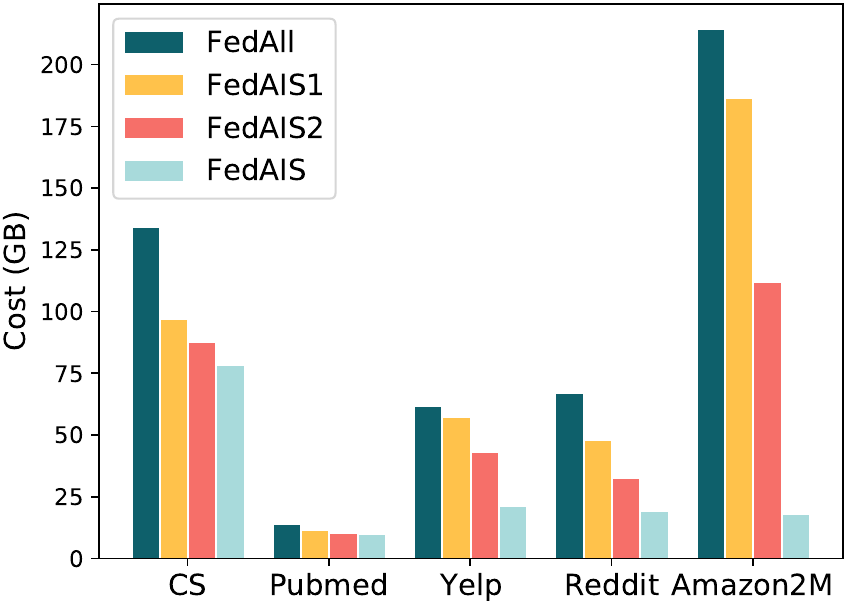}}
    	\vspace{-0.1in}
    	\caption{Model performance vs. various ablation baselines.
	}
	\label{fig:ablation} 
	    \end{minipage}
     \quad
     \begin{minipage}[t]{0.48\linewidth}
	        \subfigure[Test accuracy]{\label{fig:client_number_reddit}
	\includegraphics[width = 0.47\linewidth, height=1.18in, trim=4 4 4 4]{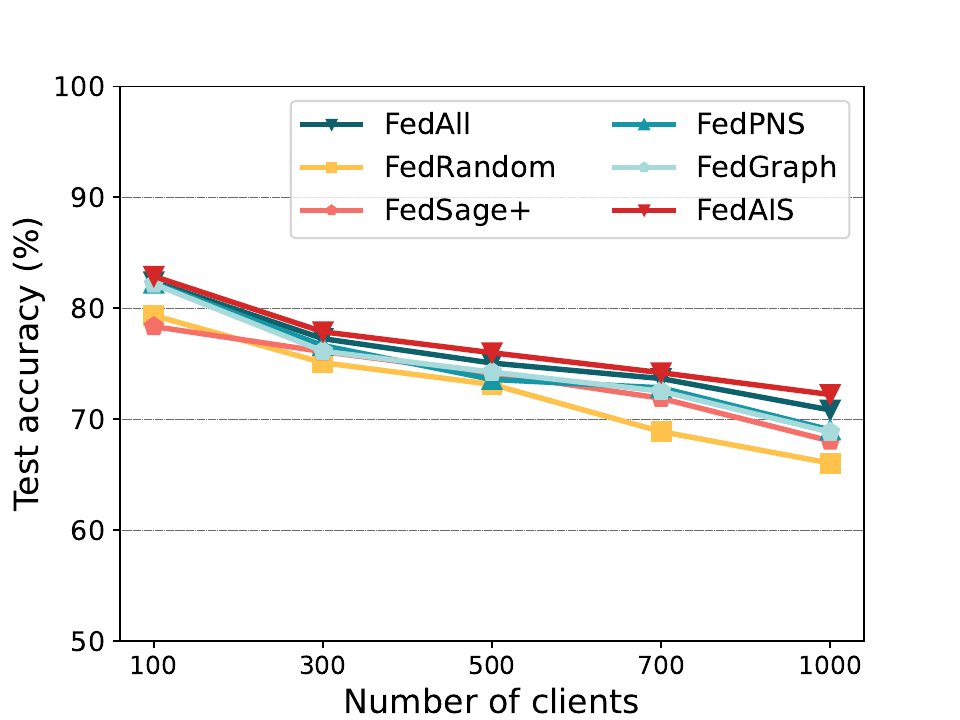}}
	\subfigure[Communication cost] {\label{fig:client_number_comm}
	\includegraphics[width = 0.47\linewidth, height=1.18in, trim=4 4 4 4]{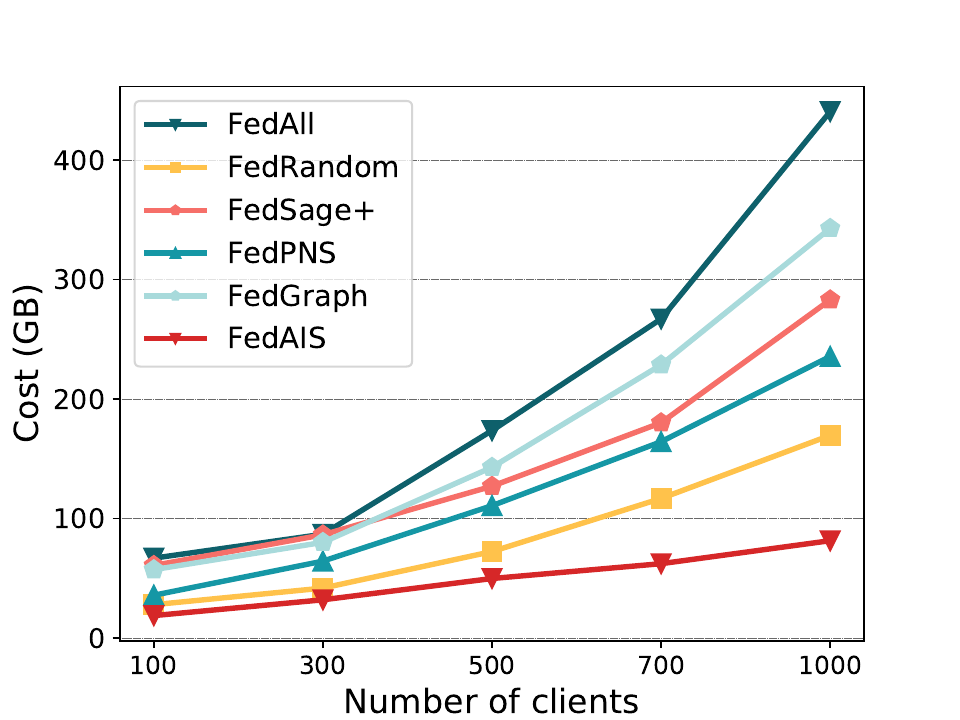}}
    	\vspace{-0.1in}
    	\caption{Model performance vs. various numbers of clients.
	}
	\label{fig:client-num} 
	    \end{minipage}
\vspace{-0.14in} 
\end{figure*}
\subsection{Results and Discussions}
\textbf{\methodname{} achieves comparable or higher accuracy.} 
We adopt three metrics \cite{falessi2021impact, herbold2018comparative}, \emph{i.e.}, test accuracy, F1-score, and area under the curve (AUC), to evaluate the accuracy of FedGCN models.
We compare \methodname{} with other baseline methods by training different FedGCN models in both iid and non-iid settings. 
We present the results of those accuracy metric scores and the standard deviations of the final global models in Table \ref{tab:perf-metric}. 
It shows that \methodname{} achieves test accuracy, F1 score, and AUC scores that are comparable to or better than other methods. For example, for model \texttt{FedYelp} trained on Yelp dataset in the iid setting, the test accuracy of \methodname{} is 4.55\%, 3.87\%, 5.20\%, 5.14 higher than other methods, respectively. 



\textbf{\methodname{} significantly saves computation and communication costs.} 
We present the test accuracy with the size of communication cost for training \texttt{FedReddit} and \texttt{FedAmazon} in iid settings in Fig. \ref{fig:testAcc_bytes}. 
The test accuracy, F1-score and AUC scores with the size of communication cost for training \texttt{FedAuthor}, \texttt{FedPubmed} and \texttt{FedYelp} in both iid and non-iid settings are much similar to that in Fig. \ref{fig:testAcc_bytes}. 
The results in Fig. \ref{fig:testAcc_bytes} show that \methodname{} requires much less amount of communication volume than the other baselines to achieve the target test accuracy, which leads to less computation time for transmitting those node embedding bytes and updating model parameters. 
Besides, we present the total computation and communication overhead for training FedGCN models in Fig. \ref{fig:comp_comm_cost}. It shows that \methodname{} achieves significantly savings of both computation and communication costs than other baselines. 
\subsection{Ablation Study}
We perform ablation studies to show the effectiveness of each component of \methodname{}. We compare \methodname{} against the following ablation baselines:
1) FedAll; 2) \texttt{FedAIS1}: it only conducts the proposed dynamic importance sampling method of local samples without adaptive embedding synchronization; 3) \texttt{FedAIS2}: it conducts training using all local samples with the proposed adaptive embedding synchronization. 
We present the test accuracy with the size of communication cost and the total communication costs in Fig. \ref{fig:ablation}. 
The results show that \methodname{}, \texttt{FedAIS1} and \texttt{FedAIS2} achieve higher performance in saving much communication costs to reach the target accuracy scores and \methodname{} performs the best among them. 
Thus, both the dynamic importance sampling module and the adaptive embedding synchronization module are effective to construct \methodname{}.  

\begin{figure}[t!]
		\centering
		\subfigure[Various non-iid degrees]{\label{fig:noniid_testAcc_reddit}
	\includegraphics[width = 0.47\linewidth, height=1.18in, trim=4 4 4 4]{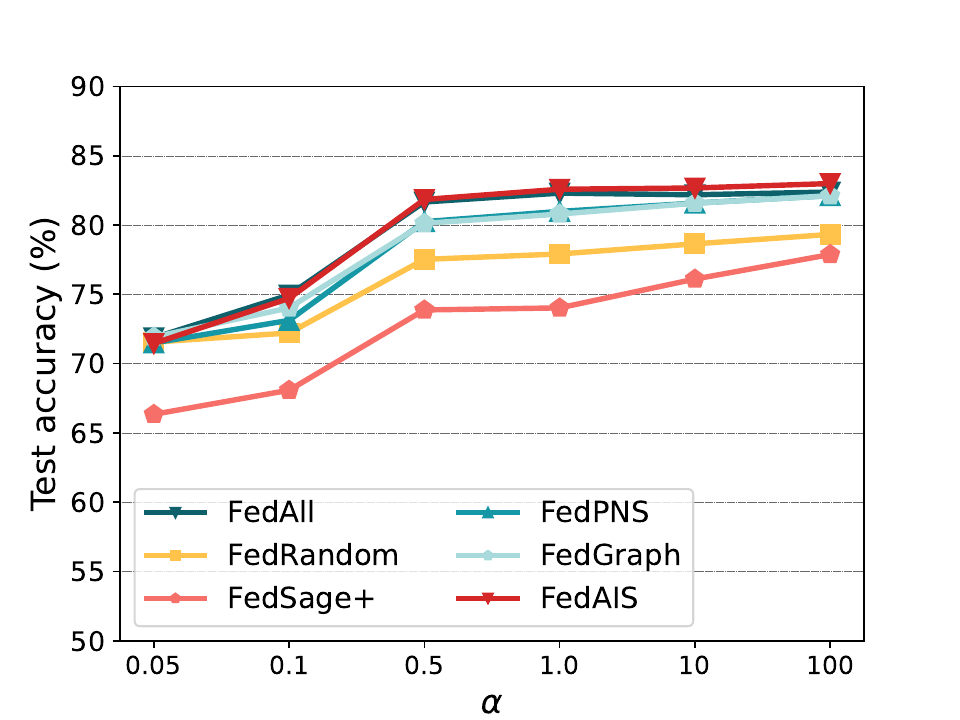}}
	\subfigure[Different sampling ratios] {\label{fig:sample_ratio_comm}
	\includegraphics[width = 0.47\linewidth, height=1.18in, trim=4 4 4 4]{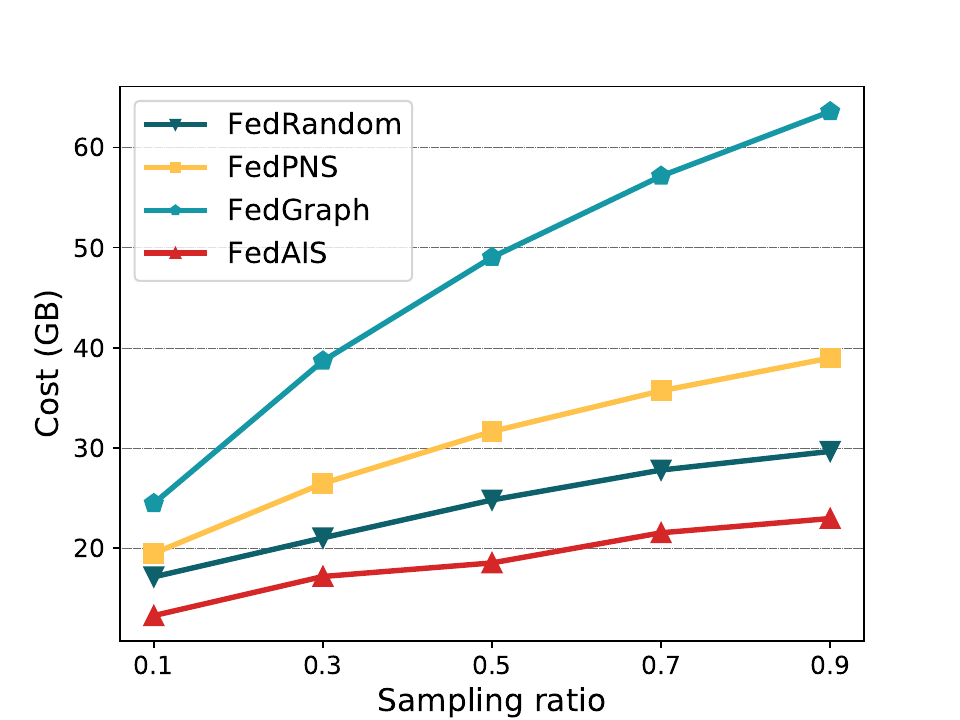}}
    	\vspace{-0.1in}
    	\caption{Sensitivity analysis of non-iid degree, sample ratio. 
	}
	\label{fig:sample_ratio}
\vspace{-0.1in} 
\end{figure}



\subsection{Sensitivity Analysis}
\label{sec:sensitivity} 
\textbf{Impact of the number of clients.} We conduct experiments with different numbers of clients engagement, \emph{i.e.}, $K=100, 300, 500, 700, 1,000$. We present the test accuracy and communication cost of the model \texttt{FedReddit} trained with different number of clients in Fig. \ref{fig:client-num}. 
It shows that the test accuracy of \methodname{} is consistently high, \emph{i.e.}, above 75.0\%, as the number of client increases to 1,000, and achieves test accuracy that is comparable to or higher than others. Besides, it shows that communication costs increase as the number of clients increases and \methodname{} achieves substantial cost savings than other baselines in all settings.

\textbf{Impact of the non-iid degree.} 
We present the test accuracy of the model \texttt{FedReddit} with different non-iid degrees, \emph{i.e.}, $\alpha=0.05, 0.1, 0.5, 1.0, 10, 100$, in Fig. \ref{fig:noniid_testAcc_reddit}. 
It shows that \methodname{} achieves accuracy scores that are comparable to or higher than other baselines. 
Besides, these accuracy scores increase as $\alpha$ increases, and when $\alpha$ is greater than 0.5, the accuracy scores are relative high. 
The accuracy scores with the sizes of communication cost for training the model \texttt{FedReddit} with different non-iid degrees are much similar to that in Fig. \ref{fig:comm_testAcc_bytesReddit}, which shows that \methodname{} consistently requires less communication cost than others.

\textbf{Impact of the ratio of samples selected.} 
Fig. \ref{fig:sample_ratio} presents the test accuracy and communication costs of the model \texttt{FedReddit} when the selection ratio is $r=$0.1, 0.3, 0.5, 0.7, 0.9. Here, we adjust FedPNS and FedGraph so that they can select corresponding ratios of nodes. 
It shows that both the test accuracy and communication cost increase as the sampling ratio increases and \methodname{} performs much better. 
Besides, due to the large size of the Reddit dataset, \methodname{} can construct a FedGCN model with high test accuracy and less communication cost by sampling only 0.1 proportion of local samples, which achieves significantly more advantageous trade-offs between accuracy and efficiency.

\section{Conclusions}
In this paper, we proposed a federated adaptive importance-based sampling approach, \methodname{}, for large-scale graph data in node classification tasks. It achieves substantial computation and communication costs by efficiently utilizing historical embedding estimators and reducing unnecessary sample training via dynamic importance-based sampling. 
Besides, it reduces cross-client neighbor embedding communication through adaptive embedding synchronization.  
In this way, \methodname{} determines the optimal communication period and achieves faster convergence with lower costs and lower prediction errors. Extensive evaluations show that \methodname{} achieves comparable or higher test accuracy, while saving significant communication and computation costs.




\bibliography{aaai24}
\end{document}